# Broadband solid cloak for underwater acoustics

Yi Chen[1], Mingye Zheng[1], Xiaoning Liu[1*], Yafeng Bi[2], Zhaoyong Sun[2], Ping Xiang[3], Jun Yang[2], Gengkai Hu[1*]

[1]Key Laboratory of Dynamics and Control of Flight Vehicle, Ministry of Education, School of Aerospace Engineering, Beijing Institute of Technology, Beijing 100081, China

[2]Key Laboratory of Noise and Vibration Research, Institute of Acoustics, Chinese Academy of Sciences, Beijing 100190, China

[3]System Engineering Research Institute, Beijing 100094, China

* Corresponding author: liuxn@bit.edu.cn, hugeng@bit.edu.cn

**Abstract**

Application of transformation theory to underwater acoustics has been a challenging task because highly anisotropic density is unachievable in water. A possible strategy is to exploit anisotropic modulus rather than density, while has not been experimentally demonstrated. We present an annular underwater acoustic cloak designed from particular graded solid microstructures. The geometry tailored microstructures mimics meta-fluid with highly anisotropic modulus through substantially suppressed shear wave. Transient wave experiments are conducted with the cloak in a designed 2D underwater waveguide system and proved excellent cloaking performance for enclosed target over broadband frequency 9–15 kHz. This finding paves the way for controlling underwater acoustics using the structured anisotropic modulus meta-fluid.

Inspired by the form invariance of Maxwell's equations under spatial mapping, a general transformation method [1, 2] was proposed to control wave propagation accurately through material distribution. Invisible cloaks as ultimate applications have been realized for electromagnetic wave [3–9], flexural plate wave [10, 11], and thermal flux [12, 13] with contemporary metamaterial [14, 15]. For acoustic cloak, meta-fluid with anisotropic density [16, 17], impossible in natural fluid, is required by transformation approach. This has been realized with perforated plate technique [18, 19] for air sound. However, cloak for underwater acoustics still faces significant challenges due to unavailable material design.

For air sound, the solid perforated plate can be treated as rigid and provide additional momentum along orthogonal direction [18–21], which induces density difference along two principal directions [18, 19]. Unfortunately, no solid material [22] can be treated as rigid in water, and the stimulated shear waves in solid will dramatically decrease the anisotropy [22, 23]. Even for air sound, effective density in two principle directions with the technique can only differ in five times or less [21], which is not sufficient large to build cylindrical cloak. Besides, this metamaterial requires fluid as working media [20–24] and is essentially fluidic with limited practical applications, such as mobility working condition.

Since the anisotropic density meta-fluid is unworkable for underwater acoustics cloak, a direct conjecture is to use meta-fluid with anisotropic modulus instead of density. Although anisotropic density meta-fluid has been investigated extensively [18–24], few researches have been concerned with realizing anisotropic modulus meta-fluid. Indeed, anisotropic modulus is a rather common property in natural solids or artificial structures. Although solids differ from fluids essentially owing to co-exist longitudinal and shear waves, it is possible to decouple the two waves and obtain fluid like materials with single bulk wave through microstructure design [25, 26], so as to separate the longitudinal and shear wave velocities stringently [27–29]. This concept coincides with long ago forwarded pentamode (PM) material [30], which was recognized recently [31] as one type of anisotropic modulus meta-fluid in regard to acoustics. Through carefully tailored geometry of PM material, its effective shear to bulk modulus ratio ($G/K$) can be smaller than 1/1000 in 3D [28] or 1/100 in 2D [25], with the two principal modulus differ in fifty times [25], which mimics strong anisotropic modulus meta-fluid. More importantly, the anisotropic modulus results from quasi-static property without resonance mechanism and works for extreme broadband range [25] if the unit cell is sufficient small. Since the modulus and density of common solids, like aluminum, copper or steel, is at the same order of magnitude as water, PM type meta-fluid is especially suitable for controlling underwater acoustics [32–36]. Although PM material has demonstrated numerically its effectiveness as anisotropic modulus fluid in underwater cloak [25], mirage [27], and metasurface applications [34], experiment validation has never been reported due to complicated microstructure design and difficulties in underwater acoustic measurements.

In this letter, we designed an annular cloak for underwater acoustics with PM material and experimentally validated its cloaking performance. Due to available strong anisotropy through microstructure design, the material parameter for cloak is not scaled as previous examples [18, 19] and is therefore impedance matched with background water. In addition, the fabricated cloak has a much smaller thickness relative to inner radius [3, 6, 10, 11], which is favorable for practical usage. Transient wave experiments were conducted in a specially designed 2D underwater waveguide system and validated superior wave shielding performance of the designed cloak, achieving average 6.3 dB reduction of target strength over a broad frequency band 9–15 kHz.

Figure 1a shows the fabricated annular cloak machined from an aluminum block using advanced electrical discharge machining (EDM) technique with high accuracy. The fabricated cloak weights 4.87 kg and differs about ten percentages from an ideal one (4.42 kg) [31]. Totally 50 sectors of cells are assembled around the $\theta$-direction, and each sector (Fig. 1b, bounded by red lines) has five graded PM unit cells along the $r$ direction. In designing the cloak, transformation theory with optimization algorithm are first used to get principal moduli $K_\theta$ and $K_r$ (Figs. 1d), and density $\rho$ for cloaking, and then the required material parameters are realized by calibrating geometry parameters (Fig. 1e) of adopted PM unit cell (Fig. 1c) (See Ref. [37]). Different from other unit cells [26], the rectangle mass components are transferred to the middle point of the thin ribs, and therefore further suppress shear wave by decreasing rigidity of the rib joints [25], which results in highly anisotropic modulus fluid behavior with single longitudinal polarized mode (See Fig. S1 in Ref. [37]). Modulus anisotropy $K_\theta/K_r$ increases from outer to inner sides, reaching 40 for the first two layers, while the outermost layer is nearly impedance matched with water ($Z_{\text{cloak}}/Z_{\text{water}}$=0.94, $Z$ represents impedance).

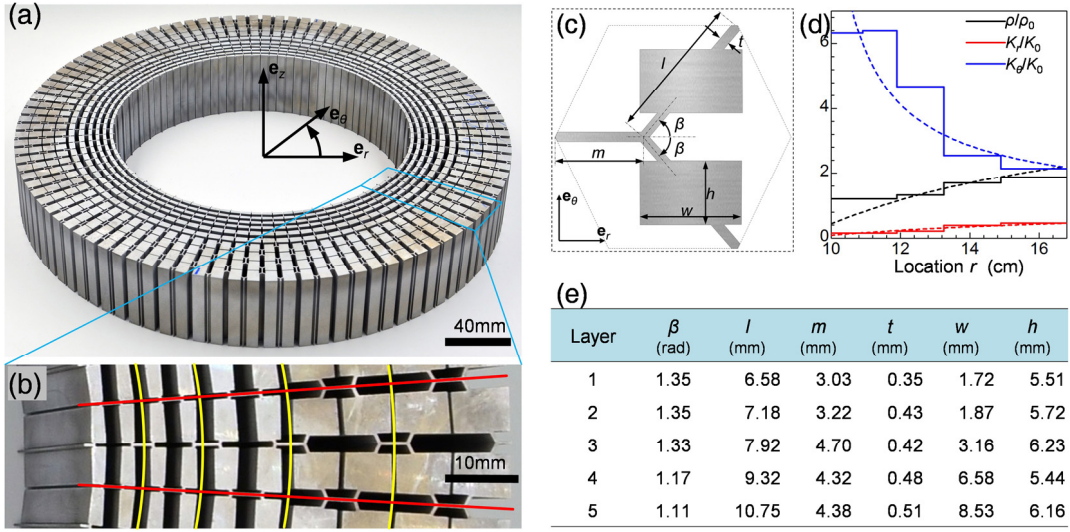

FIG. 1. Solid cloak designed from pentamode material. (a) Image of the cloak machined from an aluminum block with inner diameter 200mm, outer diameter 334 mm and height 50 mm, consisting of 500 hexagonal unit cells. (b) Enlarged view of one segment, highlighted in (a). Structure between red lines represents one sector in the $\theta$-direction and yellow lines separate the five graded unit cells. (c) Proposed pentamode unit cell with six geometric parameters, topology angle $\beta$, oblique and horizontal rib length $l$ and $m$, rib thickness $t$, rectangle width $w$ and height $h$, for tuning effective material properties. (d) Homogenized density and in-plane moduli, normalized by water density $\rho_0=1000$ kg/m$^3$ and bulk modulus $K_0=2.25$ GPa, of the designed cloak. Dashed lines represent properties derived from transformation. (e) Geometric parameters corresponding to the five graded unit cells.

Since the size of the fabricated cloak in $z$-direction is small, due to fabrication limitations, it is necessary to make measurements in a 2D underwater waveguide. However, designing 2D underwater waveguides is not trivial, since the large water density excludes any solid plate of feasible thickness from serving as an acoustic rigid boundary. Therefore, we propose a 2D underwater waveguide based on pressure compensation. The waveguide was composed of two opposing aluminum plates and a cylindrical piezoelectric transducer (PZT) immersed in an anechoic water pool (Fig. 2a). The cloak and scatter were mounted in the middle of the waveguide chamber. To keep the voids in the cloaking shell free of water, the top and bottom cloak surfaces were sealed with 4 mm thick water matched rubber. A thin walled polymethyl methacrylate (PMMA) hollow cylinder (diameter 200 mm and height 50 mm) was used as the scatter, which can be treated as air scatter with extremely small density.

Since the cloak is expected to work over broadband frequency, transient wave excitation rather than steady wave of a single frequency is preferred. We adopted a Gaussian burst of form $\exp(-0.222(f_ct)^2)\cos(2\pi f_ct)$ to drive the PZT, where $f_c$ denotes the central frequency.

Figure 2b shows the driven signal for $f_c$=13 kHz, which lasts 0.7ms to distinguish between incident and reflective signals. At this central frequency, one wavelength is approximately 11 largest unit cells of the cloak, so the homogenization condition holds. The frequency transformed signal (Fig. 2c) spans from 10–16 kHz with amplitude larger than 10% of the central frequency amplitude, hence broadband performance of the cloak can be verified. In the experiment, acoustic pressure fields corresponding to forward and backward regions (Fig. 2a) inside the waveguide for three cases (See Figs. S1 in Ref. [36]): reference (empty waveguide), uncloaked scatter, and cloaked scatter, were scanned using a hydrophone moving along orthogonal axes with 10 mm step. 3D numerical simulations have been conducted to validate the proposed waveguide with pressure compensation (See Figs. S2, S3 in Ref. [36]).

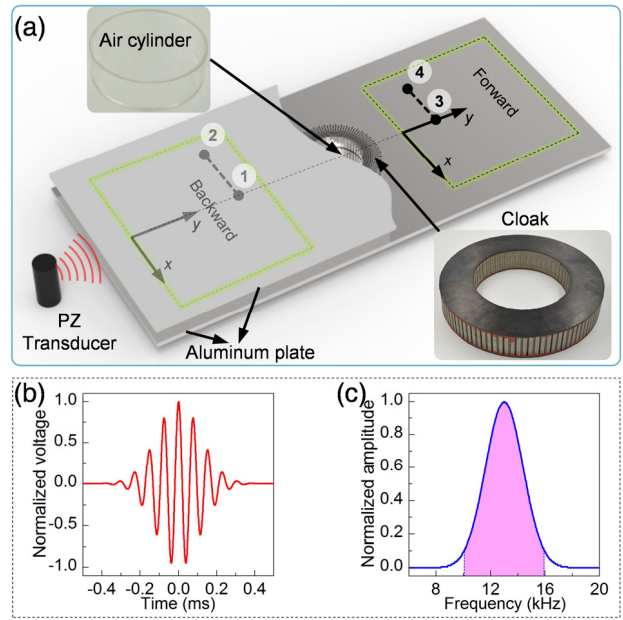

FIG. 2. Experiment setup for underwater acoustic tests. (a) 2D underwater acoustic waveguide composed of two opposing aluminum plates and transducer. Part of the upper plate is removed to show the scatter and cloak position for the experiment. Inset figures show the air cylinder used as scatter and the rubber covered cloak. Forward and backward regions (600 × 560 mm) are 450 mm apart. Locations 1 and 3 lies along the central line of the waveguide and are 475 mm away from the center, while locations 2 and 4 are 200 mm from 1 and 3, respectively. (b) (c) Time and frequency plots of Gaussian burst with central frequency 13kHz and time duration 0.7ms used for driving the transducer.

Measured snapshots of pressure fields in the forward region are shown in Fig. 3a–3c. For the reference case (Fig. 3b), pressure fields show nearly undisturbed cylindrical wave form as expected, and this provides a measurement benchmark. For the uncloaked scatter case (Fig. 3c), the impinging cylindrical wave is mostly blocked due to significant impedance mismatch with water, and a clear shadow is formed behind the scatter. A small amount of energy flows around the scatter from the lateral side, inducing phase lag owing to the longer propagation trajectory. When the scatter is covered with the cloak (Fig. 3a), impinging acoustic energy is directed by graded material properties to the forward region, and substantially eliminating the acoustic shadow to the reference case (Fig. 3b). Pressures (Fig. 3g) along the indicated horizontal line in the forward region also show very similar result for the empty waveguide and cloaked scatter cases.

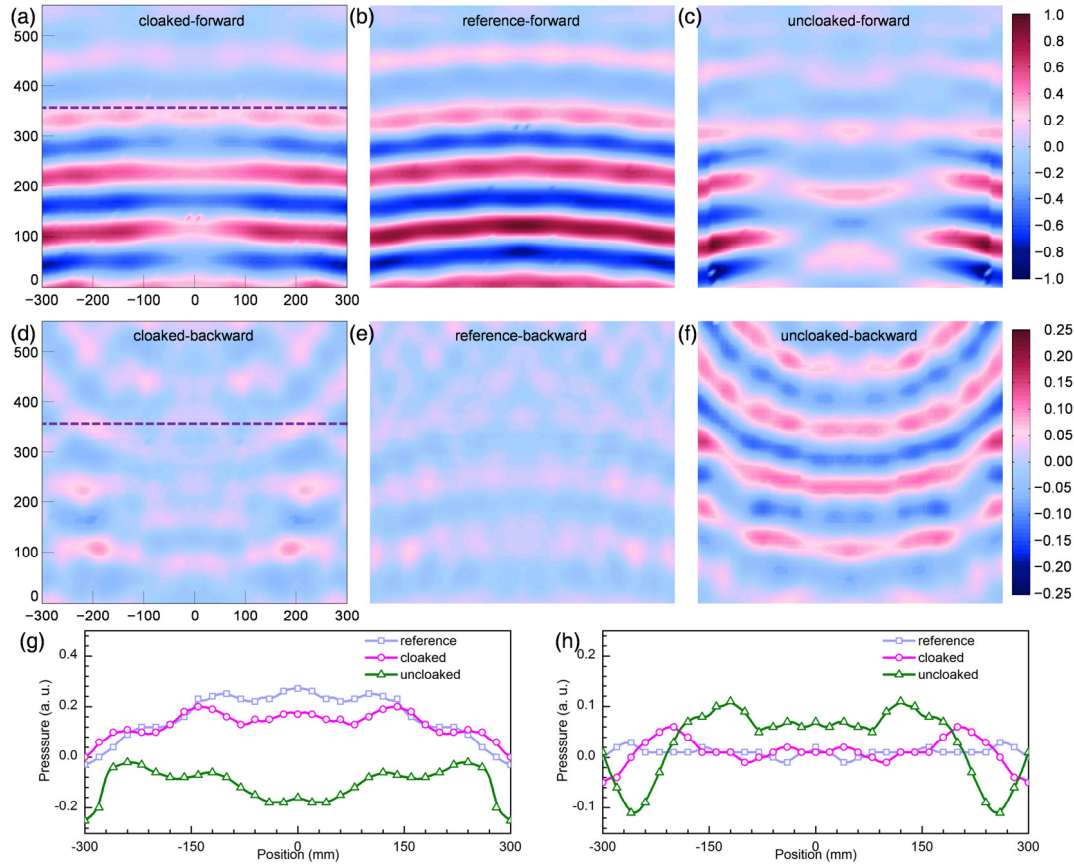

FIG. 3. Measured pressure under incidence of cylindrical Gaussian wave. (a) (b) (c) Instantaneous pressure fields in the forward region for cloaked, reference and uncloaked cases, respectively. (d) (e) (f) The same as in (a)–(c) but for the backward region. The cylindrical Gaussian wave is incident from the bottom. Pressure in the forward/backward region was normalized by largest absolute pressure of the reference case in the forward/backward region. Red/blue indicates positive/negative pressure. (g) (h) Pressure distribution for the three cases along the indicated horizontal line $y$=350 mm in forward and backward regions, respectively.

As for pressure in backward region (Fig. 3d-3e), reflected wave is only clearly observed for the uncloaked case (Fig. 3f) after the incident wave has passed through this region. The cloaked case shows much less reflection (Fig. 3d) and absolute pressure along the indicated line remains the same level as the reference case (Fig. 3h). Experimentally measured pressure fields in forward and backward regions both also agree very well with numerical simulations (See Fig. S4 in Ref. [36]). The enhanced forward transmission and reduced backward reflection can be more clearly seen from the measured transient pressure fields (see Video S1 in Ref. [36]). These results proved the impedance match of the cloak with water, and also indicate excellent cloaking effect for the enclosed scatter. Further experiment on aluminum cylinder scatter also validated the proposed underwater waveguide (see Fig. S5, Video S5 in Ref. [36]).

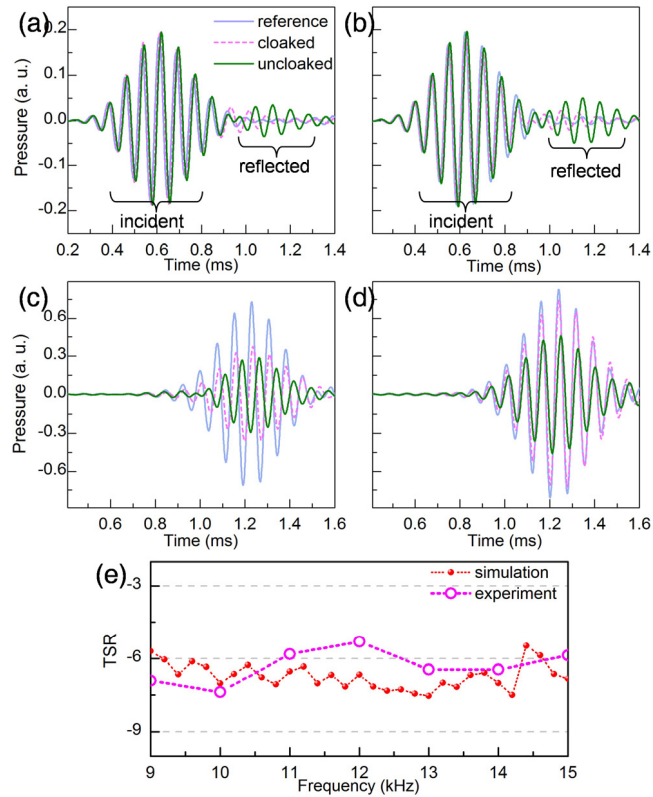

FIG. 4. Cloaking performance. (a) (b) Pressure measured at backward locations 1 and 2 from Fig. 2a. (c) (d) As (a) (b) for forward locations 3 and 4 in Fig. 2a. (e) Simulated and measured cloaked target strength reduction (TSR) for 9–15 kHz.

Figures 4a–4d show the pressures measured at locations 1–4 in Fig. 2a, respectively. At backward location 1 (Fig. 4a), the incident Gaussian packet for all three cases are similar, whereas the reflected Gaussian packet immediately following the incident packet is only clearly seen for the uncloaked case. The reflected signal partly overlaps with incident one due to large water wave velocity, and is significantly reduced when cloaked. This suppressing effect for the reflected signal is more clearly identified at another backward location (Fig. 4b). At forward location 3 (Fig. 4c), uncloaked case shows weaker and phase retarded signals as expected, and the phase retardation is rectified with moderate amplitude recovery while cloaked. The other forward location 4 (Fig. 4d) shows perfect restoration for both amplitude and phase with the cloak.

Finally, we quantified the broadband efficiency of the designed cloak using target strength reduction (TSR), calculated from the reflective signal strength. The TSR with covered cloak can be obtained as $10 \times \text{Log}_{10}(E_{cl}/E_{un})$, where $E_{cl}$ and $E_{un}$ represent energy of reflective signal for the uncloaked and cloaked cases, respectively. Additional experiments using signals with different central frequencies were conducted to cover a wide frequency band (see Videos S2–S4 in Ref. [36]). Reflective signals for both cases were taken from location 2 rather than 1 to calculate TSR with clearly distinguished reflected and incident signals. In the studied frequency range, simulated TSR (Fig. 4e) is nearly constant with small peaks caused by shear resonance scattering [24]. The overall trend of measured TSR is consistent with simulation, where differences may arise from system errors including fabrication, measurement, and imperfect waveguide. The experiment validates superior stealth capability of designed cloak, achieving 6.3 dB average reduction within a remarkably broad frequency range 9–15 kHz. It should be noted, although measurements here only indicate the excellent performance of the designed cloak over the given range owing to transducer frequency limitations, the designed cloak can in principle work from zero frequency, since only quasi-static material properties are employed [25], in contrast to cloaks based on resonance mechanisms [3, 6].

In conclusion, we have experimentally demonstrated a broadband solid cloak for underwater acoustics within designed 2D underwater waveguide. The cloak is composed of highly anisotropic pentamode solid material, which acts as anisotropic modulus meta-fluid and acquires much stronger anisotropy than any anisotropic density meta-fluid in water. The highly anisotropy in modulus enables favorable thinner cloak layer, and also provides excellent cloaking for enclosed object attributed to nearly impedance match and strongly bending capability for acoustic wave. Solidity, broadband and easily tuning feature of pentamode materials offer new possibility to control underwater acoustic waves with unprecedented flexibility. Many applications for underwater acoustic may arise from this concept, such as acoustic radiation shaping, elastic/acoustic energy transforming device, etc.

**Supplementary materials**:

Methods
Tables S1–S2
Figures S1–S5
Movies S1–S5

**Acknowledgements**

The authors would like to thank H. Jia for discussion on waveguide design, and X. Ruan for preparing part of hardware devices. Funding support from the Natural Science Foundation (Grant No. 11472044, 11221202, 11632003, 11521062) and from 111 Project (B16003) is acknowledged.

**References**

[1] J. B. Pendry, D. Schurig, and D. R. Smith, Science **312**, 1780 (2006).

[2] U. Leonhardt, Science **312**, 1777 (2006).

[3] D. Schurig, J. J. Mock, B. J. Justice, S. A. Cummer, J. B. Pendry, A. F. Starr, and D. R. Smith, Science **314**, 977 (2006).

[4] W Cai, U. K. Chettiar, A. V. Kildishev and V. M. Shalaev, Nat. Photonics **1**, 224 (2007).

[5] B. Edwards, A Alù, M. G. Silveirinha and N. Engheta, Phys. Rev. Lett. **103**, 153901 (2009).

[6] R. Liu, C. Ji, J. J. Mock, J.Y. Chin, T. J. Cui and D. R. Smith, Science **323**, 366 (2009).

[7] J. Valentine, J. Li, T. Zentgraf, G. Bartal and X. Zhang, Nat. Mater. **8**, 568 (2009).

[8] B. Zhang, Y. Luo, X. Liu and G. Barbastathis, Phys. Rev. Lett. **106**, 033901 (2011).

[9] N. Landy and D. R. Smith, Nat. Mater. **12**, 25 (2013).

[10] N. Stenger, M. Wilhelm and M. Wegener, Phys. Rev. Lett. **108**, 014301 (2012).

[11] D. Misseroni, D. J. Colquitt, A. B. Movchan, N. V. Movchan and I. S. Jones, Sci. Rep. **6**, 23929 (2016).

[12] H. Xu, X. Shi, F. Gao, H. D. Sun and B. Zhang, arXiv:1306.6835

[13] T. Han, X. Bai, D. Gao, J. T. L. Thong, B. Li and C. W. Qiu, Phys. Rev. Lett. **112**, 054302 (2014).

[14] D. R. Smith, W. J. Padilla, D. C. Vier, S. C. Nemat-Nasser and S. Schultz, Phys. Rev. Lett. **84**, 4184 (2000).

[15] Z. Liu, X. Zhang, Y. Y. Zhu, Z. Yang, C. T. Chan, and P. Sheng, Science **289**, 1734 (2000).

[16] S. A. Cummer, and D. Schurig, New J. Phys. **9**, 45 (2007).

[17] H. Chen, and C. Chan, Appl. Phys. Lett. **91**, 183518 (2007).

[18] B. I. Popa, L. Zigoneanu and S. A. Cummer, Phys. Rev. Lett. **106**, 253901 (2011).

[19] L. Zigoneanu, B. I. Popa and S. A. Cummer, Nat. Mater. **13**, 352 (2014).

[20] J. B. Pendry, and J. Li, New J. Phys. **10**, 115032 (2008).

[21] D. Torrent and J. Sanchez-Dehesa, Phys. Rev. Lett. **105**, 174301 (2010).

[22] Y. Urzhumov, F. Ghezzo, J. Hunt, and D. R. Smith, New J. Phys. **12**, 73014 (2010).

[23] B. I. Popa, W. Wang, A. Konneker, S. A. Cummer, and C. A. Rohde, J. Acoust. Soc. Am. **139**, 3325 (2016).

[24] S. Zhang, C. Xia and N. Fang, Phys. Rev. Lett. **106**, 024301 (2011).

[25] Y. Chen, X. Liu and G. Hu, Sci. Rep. **5**, 15745 (2015).

[26] C. N. Layman, C. J. Naify, T. P. Martin, D. C. Calvo and G. J. Orris, Phys. Rev. Lett. **111**, 024302 (2013).

[27] A. C. Hladky-Hennion, J. O. Vasseur, G. Haw, C. Croënne, L. Haumesser and A. N. Norris, Appl. Phys. Lett. **102**, 144103 (2013).

[28] A. Martin, M. Kadic, R. Schittny, T. Bückmann and M. Wegener, Phys. Rev. B **86**, 155116 (2012).

[29] M. Kadic, T. Bückmann, N. Stenger, M. Thiel and M. Wegener, Appl. Phys. Lett. **100**, 191901 (2012).

[30] G. W. Milton and A. V. Cherkaev, J. Eng. Mater-T. Asme **117**, 483 (1995).

[31] A. N. Norris, P. Roy. Soc. A-Math. Phy. **464**, 2411 (2008).
[32] C. L. Scandrett, J. E. Boisvert and T. R. Howarth, J. Acoust. Soc. Am. **127**, 2856 (2010).
[33] J. Cipolla, N. Gokhale, A. N. Norris and A. Nagy, J. Acoust. Soc. Am. **130**, 2332 (2011).
[34] Y. Tian, Q. Wei, Y. Cheng, Z. Xu, Z. and X. Liu, Appl. Phys. Lett. **107**, 221906 (2015).
[35] X. Cai, L. Wang, Z. Zhao, A. Zhao, X. Zhang, T. Wu and H. Chen, Appl. Phys. Lett. **109**, 131904 (2016).
[36] Y. Chen, X. Liu and G. Hu, J. Acoust. Soc. Am. **140**, EL405 (2016).
[37] See Supplemental Material for detail on cloak design and experiment setup, additional analysis on proposed underwater waveguide, calculations and videos, which includes Ref. [38–41].
[38] N. H. Gokhale, J. L. Cipolla and A. N. Norris, J. Acoust. Soc. Am. **132**, 2932 (2012).
[39] A. S. Titovich and A.N. Norris, J. Acoust. Soc. Am. **136**, 1601 (2014).
[40] W. Chen, Z. Bian and H. Ding, Int. J. Mech. Sci. **46**, 159 (2004).
[41] D. N. MacLennan and E. J. Simmonds, *Fisheries Acoustics* (Springer, 2013).